# Contact Formation and Viscoelastic Detachment in Non-Circular Soft Adhesive Contacts

Sonu Dhiman, Debashish Das*

Mechanical Engineering, Indian Institute of Science, Bengaluru-560012, India

## ABSTRACT

Adhesive contact measurements on soft polymers are commonly interpreted using Johnson-Kendall-Roberts (JKR) theory, which is formulated for circular contacts. Non-circular contacts generally require more complex elliptical-contact descriptions, making it unclear when a scalar measure of contact size is sufficient. Here, we examine contact formation and detachment in non-circular soft adhesive contacts using PDMS crossed-cylinder experiments. The cylinder crossing angle was varied from 30º to 90º, producing contacts ranging from highly elongated ellipses to nearly circular geometries while keeping the material pair fixed. During loading, the contact aspect ratio $b/a$ initially evolved but rapidly approached an angle-dependent plateau, indicating approximately self-similar growth. This motivates the use of the area-equivalent radius $c = \sqrt{ab}$ together with the geometric-mean curvature radius $R_e = \sqrt{R_1 R_2}$. With these substitutions, the loading branches follow a JKR-type linearization and yield a nearly angle- and preload-independent loading work of adhesion, $W_{\text{load}}$ = 24 ± 1 mJm$^{-2}$. Comparable values obtained from Johnson–Greenwood elliptical-contact fits confirm that this result is not an artifact of neglecting ellipticity. In contrast, unloading and pull-off are strongly history dependent. The unloading branches can still be represented by an effective JKR-like relation using $c = \sqrt{ab}$, but only with a substantially larger effective separation energy, $W_{\text{unload,eff}}$, which increases with preload and with decreasing crossing angle. A reduced viscoelastic model based on the same area-equivalent description further captures the principal unloading response over 50º-80º using a single shared parameter set across different angles and preloads. These results show that contact formation in elliptical soft adhesive contacts is governed primarily by area-equivalent scaling, whereas detachment is governed by geometry- and history-dependent dissipative separation. The framework provides a practical route for analyzing non-circular soft contacts using simple global measures while retaining the essential distinction between contact formation and viscoelastic detachment.

***Keywords:*** JKR theory; elliptical contact; crossed cylinders; PDMS; viscoelasticity; work of adhesion; preload dependence

* Corresponding Author: Email: ddas@iisc.ac.in

## 1. Introduction

Adhesion between compliant solids is governed by a competition between elastic deformation and interfacial energy. This coupling is central to contact mechanics of elastomers, gels, pressure-sensitive adhesives, biological tissues, soft robotic grippers, microstructured adhesives, and tribological interfaces. For soft materials, even modest surface forces can produce finite contact areas, making contact geometry an experimentally accessible route to determine elastic and adhesive properties. As a result, contact mechanics has become one of the standard approaches for measuring the work of adhesion and for studying the mechanics of soft interfaces [1–7].

The classical Johnson-Kendall-Roberts (JKR) theory [1] provides the foundational description of adhesive elastic contact in the limit of compliant solids, short-range adhesion, and relatively large contact radii. In this framework, adhesion is incorporated through an energy balance analogous to linear elastic fracture mechanics, leading to direct relationships between load, contact radius, elastic modulus, radius of curvature, and work of adhesion. The JKR solution is especially powerful because, for axisymmetric contacts such as a sphere on a flat surface or two spheres in contact, a single scalar contact radius completely describes the contact size. The theory also predicts a tensile pull-off instability, from which the work of adhesion can be estimated using the effective radius of curvature. Related theories, including the Derjaguin-Muller-Toporov [8] and Maugis-Dugdale [9] models, describe other regimes depending on the range of adhesive forces, elastic compliance, and contact length scale.

Despite its success, the standard JKR formulation is geometrically restrictive. Many contacts of practical interest are not circular. Cylindrical lenses, fibers, anisotropic asperities, textured surfaces, and non-orthogonal crossed cylinders can form line-like, elliptical, or otherwise non-axisymmetric contact regions. In such cases, the contact boundary cannot generally be represented by a single circular crack front. The local stress intensity factor or energy-release rate may vary around the perimeter, and the contact shape may evolve with load. This makes the interpretation of non-circular adhesive contacts substantially more difficult than that of axisymmetric JKR contacts.

Cylindrical and crossed-cylinder geometries have a long history in measurements of surface forces and adhesion. In particular, the surface forces apparatus (SFA) employs two smooth surfaces mounted in a crossed-cylinder configuration to quantify normal surface forces, adhesion, and interfacial interactions with high resolution [5, 10, 11]. The surfaces may be bare mica or coated with polymers, surfactants, or other thin films, allowing the adhesion and friction of polymeric interfaces to be investigated directly [12, 13]. In the conventional orthogonal crossed-cylinder configuration, equal-radius cylinders produce a locally axisymmetric gap. More generally, however, varying the angle between the cylinders introduces unequal principal curvatures and provides a controlled transition from circular to increasingly elliptical contacts [14]. This makes the crossed-cylinder geometry particularly useful for examining how contact shape influences adhesive contact formation and detachment.

Theoretical treatments of elliptical adhesive contacts are more complex than circular JKR theory. Johnson and Greenwood developed an approximate JKR theory for elliptical contacts by extending the fracture-mechanics interpretation of adhesion to a non-axisymmetric contact boundary [15]. In this type of formulation, the contact eccentricity, load, and contact dimensions are coupled, and the contact shape is not generally expected to remain geometrically similar during loading. More recent theoretical and numerical studies have revisited elliptical JKR-type contacts and have shown that the accuracy of approximate elliptical theories can depend strongly on the degree of eccentricity [16, 17]. These developments highlight an important unresolved experimental issue: although elliptical adhesive contacts can be treated with detailed non-axisymmetric theories, it remains unclear when such complexity is required for interpreting measured contact data and when a simpler effective description is sufficient.

This issue is particularly important for soft polymers such as polydimethylsiloxane (PDMS). PDMS is widely used in adhesion and contact-mechanics experiments because it is compliant, optically accessible, chemically tunable, and capable of forming measurable adhesive contacts. However, PDMS contacts are rarely ideal elastic interfaces. Even when the bulk response of PDMS is approximately elastic, adhesive contacts can exhibit pronounced loading–unloading hysteresis due to rate-dependent viscoelastic dissipation and interfacial processes. Previous studies have demonstrated sensitivity to contact time, network and dangling-chain structure, surface chemistry, humidity, and substrate surface roughness [18–22]. Consequently, the work of adhesion inferred from contact formation need not coincide with the effective separation energy inferred from unloading or pull-off. This distinction becomes especially important when comparing contact growth, crack-front recession, and terminal detachment within the same interface.

A related open question is how contact geometry influences these two stages differently. During loading, if dissipative effects are limited, the measured adhesion should be relatively insensitive to contact shape provided that an appropriate scalar measure of contact size and curvature is used. For an elliptical contact with semi-axes $a$and $b$, the area-equivalent radius $c = \sqrt{ab}$ preserves the contact area through $A = \pi ab = \pi c^2$. Likewise, for a non-axisymmetric quadratic gap, the geometric-mean radius $R_e = \sqrt{R_1 R_2}$ provides a natural scalar curvature measure. These quantities suggest that the global loading response of an elliptical contact may admit an effective circular JKR representation. Whether such a reduction remains useful during unloading is less obvious, because detachment is history dependent and the local crack-front motion may become strongly nonuniform.

Here, we investigate adhesive PDMS contacts using sphere-based and crossed-cylinder geometries, with particular emphasis on non-orthogonal crossed cylinders. The crossed-cylinder configuration allows the contact footprint to be varied continuously from nearly circular to highly elongated elliptical while keeping the material pair fixed. We first test whether the loading response can be reduced to an area-equivalent JKR description using $c = \sqrt{ab}$ and $R_e = \sqrt{R_1 R_2}$, and compare the resulting adhesion values with Johnson–Greenwood elliptical-contact fits. We then examine unloading and pull-off to determine how preload and contact elongation influence the effective separation response. Finally, we ask whether the non-circular unloading trajectories can be described by a reduced viscoelastic formulation based on the same area-equivalent coordinate, and whether a common set of viscoelastic parameters can represent contacts spanning different crossing angles and preloads. In this way, the study separates the comparatively geometry-insensitive mechanics of contact formation from the geometry- and history-dependent mechanics of detachment.

The present study therefore advances beyond existing elliptical-contact analyses in three respects: it experimentally validates an area-equivalent JKR reduction for non-circular loading, identifies a systematic geometry-preload coupling during detachment, and demonstrates that the resulting unloading trajectories can be described using a shared-parameter reduced viscoelastic model. Together, these results provide a unified framework for interpreting both formation and separation of soft non-circular adhesive contacts using experimentally accessible global contact measures.

## 2. Materials and Experimental Methods

### *2.1. Specimen (PDMS) Preparation along with Contact Geometries*

PDMS specimens were prepared from Sylgard 184 (Dow), Fig. 1, using a base-to-curing-agent ratio of 10:1 by weight. The uncured mixture, after degassing to remove any trapped air bubbles, was poured into aluminum molds to form a 7 mm-thick base layer with an approximately 2 mm-high raised contact feature. The surface geometry of the contact feature was defined by molding the PDMS against precision optical lenses. Spherical contacts were produced using plano-concave lenses of radius $R = 9.42$ mm, while cylindrical contacts were produced using plano-cylindrical-concave lenses of radius $R = 10.33$ mm. Flat surfaces were obtained from the planar mold/lens surface. The molded geometries provided smooth PDMS interfaces for sphere-on-sphere, sphere-on-flat, and crossed-cylinder contact experiments. After casting, the PDMS was cured at 100°C for 2h and oven cooled to room temperature and then demolded to produce the final specimen.

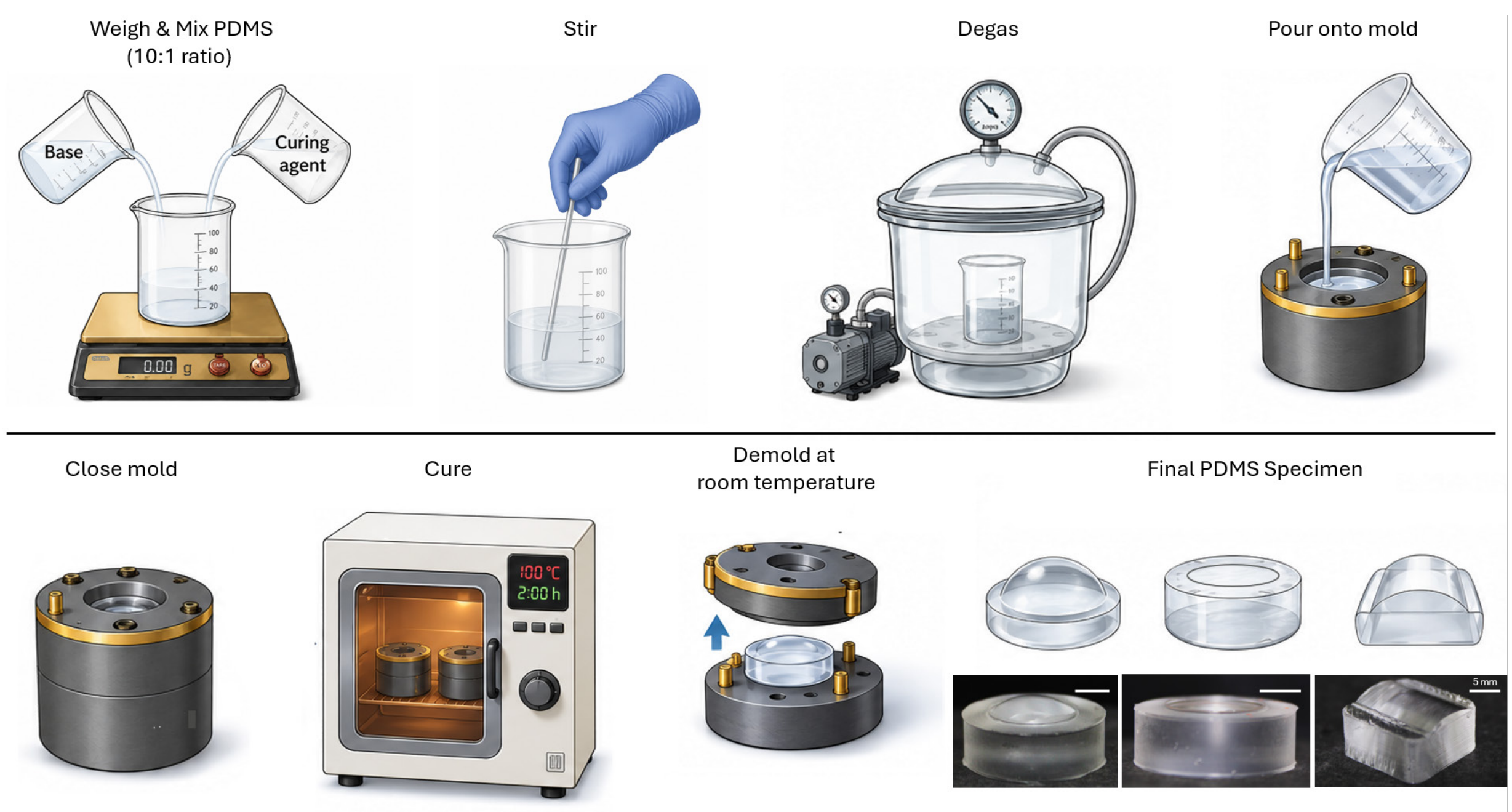

**Fig. 1.** Schematic of the fabrication process to create PDMS contact surfaces

### *2.2. Experimental Setup*

Contact experiments were performed in a clean laboratory environment at 26 ± 2 ºC and 70 ± 5% relative humidity. The experimental setup is shown in Fig. 2(a). A compact motorized XYZ translation stage was used to control the relative displacement between the two PDMS specimens. A manual precision micrometer rotation stage was mounted on the XYZ stage to vary the crossing angle between the cylindrical contact surfaces in the cylinder-cylinder experiments.

The normal force was measured using a 100 g load cell with 1 μN load resolution. The load cell was mounted directly on the rotation stage, and the lower PDMS specimen was fixed to the load cell through a metal load plate using a thin layer of adhesive. The upper PDMS specimen was bonded to a glass microscope slide and mounted above the lower specimen. The upper mounting plate contained an optical access window, allowing the contact interface to be imaged, *in situ* using an optical microscope, from above during loading, dwell, and unloading (see. Fig. 2(b)).

### *2.3. Experimental Protocols*

All samples were stored in a closed chamber after fabrication and tested one day later. Before each experiment, the two PDMS specimens were aligned laterally. For sphere-on-sphere (S-o-S) and cylinder-on-cylinder (C-o-C) contacts, alignment in the $x$- and $y$-directions was performed using side-view shadowgraphy. This step was particularly important for S-o-S contacts, where the two curved surfaces must meet near their apexes, and for C-o-C contacts, where the intended crossing configuration must be maintained. For sphere-on-flat (S-o-F) contacts, lateral alignment was less restrictive because the flat substrate does not impose a unique apex contact point.

After alignment, the setup was placed under an optical microscope (Olympus BX53). The contact interface was imaged from above using a 2.3 MP resolution camera at 4 frames/s. During each test, the lower specimen was displaced in the $z$-direction at a constant rate of 1 μm/s until the prescribed preload was reached. Four nominal preloads were used: 0.025, 0.050, 0.075, and 0.100 N. Upon reaching the target preload, the stage motion was stopped, and its position was held fixed for 10 s. No active force control was applied during this dwell period; both the normal force and contact dimensions were therefore free to evolve at fixed imposed displacement. After the dwell, the stage direction was reversed, and the specimen was retracted at the same displacement rate until complete detachment occurred.

The normal force was recorded using a LabVIEW-based data-acquisition system. The stage motion was controlled through LabVIEW by calling a Python script that used the measured load only as a trigger to terminate the approach at the prescribed preload and initiate the fixed-position dwell and subsequent retraction sequence.

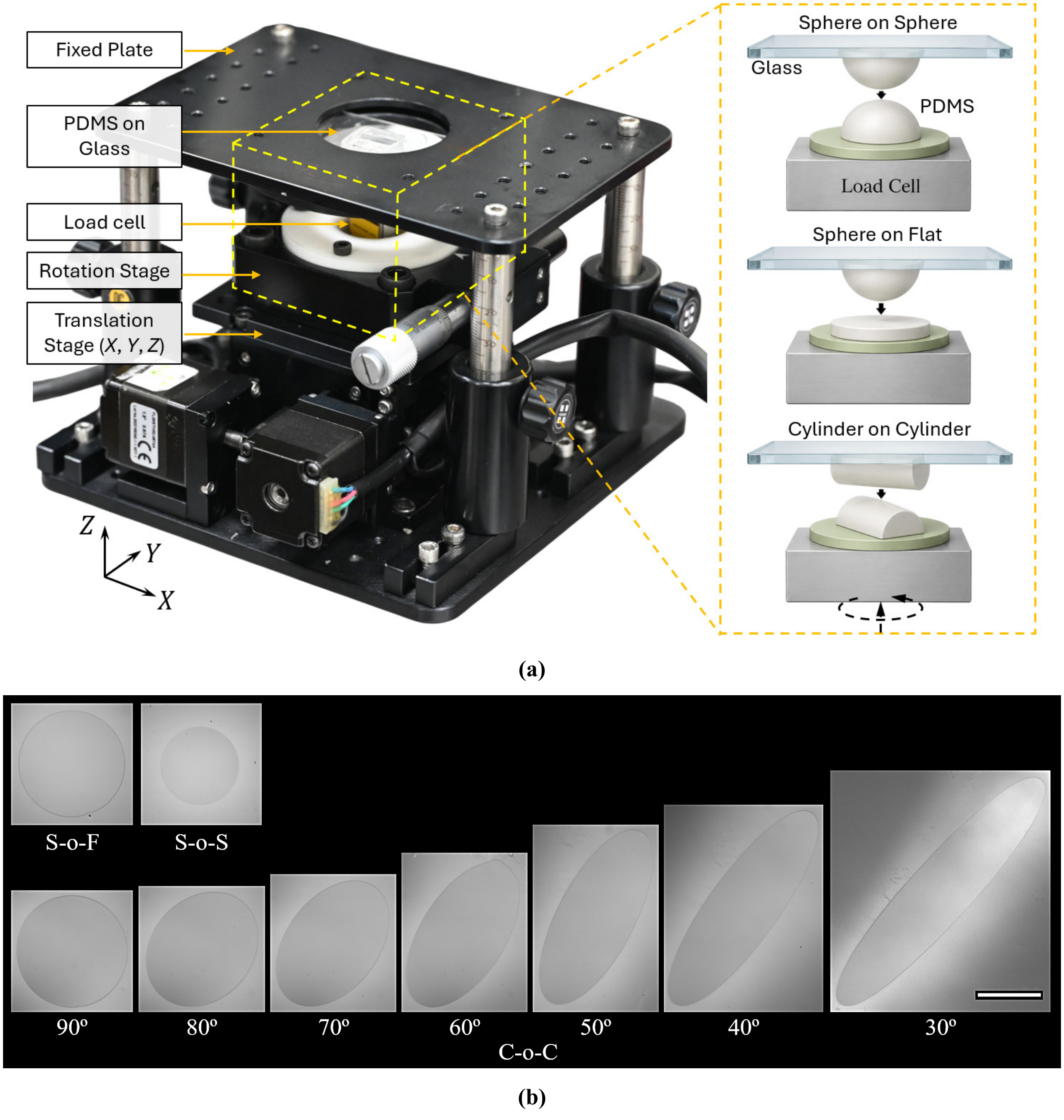

**Fig. 2. (a)** Experimental setup used to perform adhesive contact measurements between PDMS surfaces. The lower PDMS specimen was mounted on a load cell attached to a rotation stage, allowing the crossing angle between cylindrical specimens to be varied precisely. The upper PDMS specimen was mounted on a glass slide, and the contact interface was imaged optically from above through the access window. **(b)** Representative contact shapes at the end of the 10 s dwell period for a preload of 0.025 N. The images show the different contact

geometries investigated in this work, including sphere-on-flat (S-o-F), sphere-on-sphere (S-o-S), and cylinder-on-cylinder (C-o-C) contacts at different crossing angles. The scale bar is 1 mm and is common to all images.

The recorded images were first processed using Gaussian filtering and automatic contrast adjustment to reduce high-frequency image noise and improve the distinction between the contacting and non-contacting regions. The contact boundary was then identified using Canny edge detection, and a best-fit circle or ellipse was applied to the extracted perimeter. For circular contacts, the contact radius $r$ was extracted. For elliptical contacts, the semi-major and semi-minor axes, $a$ and $b$, were obtained, and the area-equivalent contact radius was calculated. The force data were synchronized with the image-based contact measurements. All subsequent contact mechanics analysis was performed using MATLAB.

## 3. Contact-Mechanics Framework

### *3.1. Geometry of crossed-cylinder contacts*

For two cylinders of equal radius $R$ crossed at an angle $\phi$, the local undeformed gap near the center of contact can be approximated by a quadratic form,

$$h(x, y) = \frac{x^2}{2R_1} + \frac{y^2}{2R_2} \tag{1}$$

where $R_1$ and $R_2$ are the principal radii of curvature of the relative gap. For equal-radius crossed cylinders, these radii are

$$R_1 = \frac{R}{1 - \cos\phi}, \ R_2 = \frac{R}{1 + \cos\phi} \tag{2}$$

with $R_1 \geq R_2$for $\phi \leq 90°$. A useful scalar measure of the non-axisymmetric curvature is the geometric-mean radius,

$$R_{\text{eff}} = \sqrt{R_1 R_2} = \frac{R}{\sin\phi} \tag{3}$$

At $\phi = 90°$, $R_1 = R_2 = R$, and the local gap becomes axisymmetric. The corresponding contact is circular and is equivalent to the classical circular JKR geometry with effective radius $R_{\text{eff}} = R$. For $\phi < 90°$, the principal radii are unequal and the contact becomes elliptical. Thus, by varying the crossing angle, the contact shape can be changed continuously from circular to highly elongated while keeping the material pair and surface preparation unchanged.

The ellipticity of the contact is characterized by the ratio $g = b/a,$ where $a$ and $b$ are the semi-major and semi-minor axes of the measured contact region, respectively. The corresponding area-equivalent contact radius is defined as

$$c = \sqrt{ab} \tag{4}$$

so that the measured elliptical contact area is preserved:

$$A = \pi ab = \pi c^2 \tag{5}$$

Therefore, $c$ is the radius of a circular contact having the same area as the measured elliptical contact. In the analysis below, $c$ is used as the scalar measure of contact size for non-circular contacts.

### *3.2. Area-equivalent JKR representation*

For a circular adhesive contact, the JKR load - contact radius relation can be written as

$$P = K\frac{a^3}{R} - \sqrt{6\pi K W a^3}, \qquad K = \frac{4E^*}{3} \tag{6}$$

where $P$ is the normal load, $E^*$ is the reduced elastic modulus, $W$ is the work of adhesion, and $R$ is the effective radius of curvature. The applicability of the JKR regime was assessed using the Tabor parameter, $\mu_T = [R_{\text{eff}} W^2/(E^{*2} Z_0^3)]^{1/3}$. Taking 0.3 nm as a characteristic interaction range, and using the measured elastic modulus and loading work of adhesion for PDMS, the Tabor parameter is found to be of order $10^3$ across the investigated geometries, confirming that the contacts lie well within the JKR limit.

For the non-circular crossed-cylinder contacts, we examine whether the same global form can describe the loading response when the circular contact radius and curvature radius are replaced by their area-equivalent counterparts, $c = \sqrt{ab}$ and $R_{\mathrm{eff}} = \sqrt{R_1 R_2}$. This gives

$$P = K\frac{c^3}{R_{\mathrm{eff}}} - \sqrt{6\pi K W_{\mathrm{load}} c^3} \tag{7}$$

Here, $W_{\mathrm{load}}$ denotes the effective work of adhesion associated with contact formation during loading. This expression is used as an experimentally testable area-equivalent reduction of the non-circular contact problem. It does not imply that the elliptical contact edge is locally equivalent to a circular JKR crack front. Rather, it tests whether the measured loading response is governed primarily by the evolution of the total contact area.

The corresponding linearization [23, 24] is:

$$\frac{c^{3/2}}{R_{\mathrm{eff}}} = \frac{1}{K}\frac{P}{c^{3/2}} + \sqrt{\frac{6\pi W_{\mathrm{load}}}{K}} \tag{8}$$

Therefore, a linear fit of $c^{3/2}/R_{\mathrm{eff}}$ versus $\mathrm{P}/c^{3/2}$ gives

$$K = \frac{1}{\mathrm{slope}},\ E^* = \frac{3K}{4},\ W_{\mathrm{load}} = \frac{K(\mathrm{intercept})^2}{6\pi} \tag{9}$$

The area-equivalent representation is expected to be most appropriate when the contact evolves approximately self-similarly; that is, when changes in contact area dominate the response while the aspect ratio $g = b/a$ varies only weakly over the fitted loading range. This condition is not imposed a priori. It is evaluated experimentally in the Results section by tracking the evolution of $g$ with load for different crossings.

### *3.3. Johnson–Greenwood elliptical-contact comparison*

Because the crossed-cylinder contacts are elliptical for $\phi < 90°$, the area-equivalent JKR representation is compared with the Johnson–Greenwood approximate theory for elliptical adhesive contacts [15]. In the Johnson–Greenwood formulation, the contact aspect ratio $g = b/a$, the mean contact radius $c$, and the applied load are coupled through the non-axisymmetric contact geometry. The Johnson–Greenwood model therefore serves as a more explicit elliptical-contact benchmark against which the area-equivalent reduction can be assessed.

In the present analysis, the Johnson–Greenwood model is used through global fitting of the measured $c$-versus-$P$ trajectories. Pointwise inversion of individual $(P, a, b)$ measurements is highly sensitive to the measured value of $g$. Small errors in contact-boundary detection or ellipse fitting can produce large fluctuations in $g$, especially near the circular limit where the distinction between major and minor axes becomes poorly conditioned. Global fitting provides a more stable basis for comparing the area-equivalent JKR and Johnson–Greenwood descriptions.

The comparison is used to determine whether the simpler area-equivalent representation gives loading adhesion values consistent with an explicit elliptical-contact model. Agreement between the two analyses would indicate that the global loading response is governed primarily by the area-equivalent contact size rather than by detailed pointwise variations in the instantaneous ellipse ratio.

### *3.4. Treatment of loading and unloading branches*

The loading and unloading branches were analyzed separately because they represent physically distinct interfacial processes. During loading, the contact edge advances and the extracted energy scale is denoted $W_{\mathrm{load}}$. This quantity characterizes contact formation under the prescribed loading rate, dwell history, and environmental conditions. For circular contacts, $W_{\mathrm{load}}$ was obtained from the JKR linearization, while for elliptical crossed-cylinder contacts it was obtained using the area-equivalent variables $c$ and $R_{\mathrm{eff}}$ and compared with global Johnson-Greenwood model fits.

Unloading was not treated as the reversible continuation of loading. Immediately after load reversal, the applied force decreased while the contact size changed only weakly, indicating a delayed-recession or approximately constant-area regime. Because the contact edge is not yet undergoing sustained recession, this initial stage cannot be represented by a quasi-static JKR relation or by the elastic Johnson-Greenwood model. Once the contact begins to shrink progressively, the receding portion of the unloading branch can be described over a finite interval using an effective separation energy, $W_{\text{unload,eff}}$. This quantity is not interpreted as the thermodynamic work of adhesion; rather, it represents the combined resistance arising from intrinsic interfacial adhesion, viscoelastic deformation, interfacial rearrangement, and nonuniform contact-edge recession.

Accordingly, the two fitted energy scales are interpreted as $W_{\text{load}}$: effective contact-formation energy and $W_{\text{unload,eff}}$: effective dissipative separation energy. The separate treatment of loading and unloading reflects the physical distinction between advancing and receding contact edges. The loading branch characterizes contact formation, whereas the unloading branch captures the history- and geometry-dependent resistance to separation. Section 3.5 introduces the viscoelastic framework used to interpret the delayed contact recession after load reversal and the subsequent evolution of the apparent separation energy.

### *3.5. Viscoelastic contact representation*

The elastic JKR framework provides a useful description of equilibrium adhesive contacts, but it does not account explicitly for the rate-dependent resistance to crack recession observed during unloading of viscoelastic solids. In the present experiments, the contact was formed at a displacement rate of $1\ \mu\text{m s}^{-1}$ and subsequently held for $10$ s at the maximum load. Both the normal force and contact size reached near-stationary plateau within approximately 10 s and exhibited little subsequent evolution (Fig. S1). Therefore, the contact state at the end of the 10 s dwell was taken as an approximately relaxed initial condition for the unloading analysis, as also in Ref. [7]. Accordingly, the viscoelastic model was applied only to the unloading branch, rather than to the complete loading-unloading cycle.

For a circular adhesive contact of radius $a$, the apparent work required to move the contact edge can be obtained by rearranging the JKR load-radius relation. Defining the Hertzian load corresponding to the instantaneous contact radius as [7, 25]:

$$P_H = \frac{4E_\infty^* a^3}{3R} \tag{10}$$

the apparent separation energy is written as

$$W_{\text{app}} = \frac{(P_H - P)^2}{8\pi E_\infty^* a^3} \tag{11}$$

where $P$ is the measured normal load, $R$ is the effective radius of curvature, and $E_\infty^*$ is the relaxed contact modulus. Normalization by the intrinsic work of adhesion $W$ gives the dimensionless crack-driving parameter:

$$\chi = \frac{W_{\text{app}}}{W} = \frac{(P_H - P)^2}{8\pi E_\infty^* W a^3} \tag{12}$$

For a circular contact, the crack front is axisymmetric and every point along the perimeter is mechanically equivalent. The crack-edge recession speed can therefore be represented by the single scalar quantity

$$V = -\frac{da}{dt} \tag{13}$$

Following the Greenwood-Johnson viscoelastic crack-growth formulation [7, 25], the recession speed is related to $\chi$ through:

$$-\frac{da}{dt} = C_v \frac{\chi}{\ln\left[\frac{1-k}{1-1/\chi}\right]} \tag{14}$$

where $k = E_\infty^*/E_0^*$, with $E_0^*$ and $E_\infty^*$ denoting the instantaneous and relaxed moduli, respectively. The velocity scale $C_v$ may be written as

$$C_v = \frac{\pi E_\infty^* h_0^2}{12 W \tau} \tag{15}$$

where $h_0$ is a characteristic cohesive-zone length and $\tau$ is a viscoelastic relaxation time. The admissible range of the constitutive relation is $1 < \chi < \frac{1}{k}$. As $\chi$ approaches $1/k$, the denominator of the recession law tends to zero and the predicted crack speed increases sharply, indicating the approach to terminal instability. For the circular contacts, the relaxed contact modulus $E_\infty^*$ and intrinsic work of adhesion $W$ were obtained from the slow loading response using the elastic JKR framework, while the modulus ratio $k = E_\infty^*/E_0^*$ and circular-contact velocity scale $C_v$ were identified by fitting the viscoelastic model to the unloading branch.

For crossed-cylinder contacts, the footprint is elliptical and is characterized by two semi-axes, $a$ and $b$. A complete viscoelastic treatment would require the local energy-release rate and local crack-edge velocity to be resolved around the evolving elliptical perimeter. The experiments nevertheless show that, for crossing angles between $50°$and $80°$, the aspect ratio varies only weakly over most of unloading up to the maximum tensile load. The contact therefore evolves approximately as a one-parameter family of geometrically similar ellipses over this interval. This motivates the use of the area-equivalent coordinate $c$ as a reduced global contact-size variable.

The corresponding area-equivalent Hertzian load is defined as

$$P_H(c) = \frac{4E_\infty^* c^3}{3R_{\text{eff}}} \tag{16}$$

and the apparent energy ratio inferred from the measured $P - c$ response is

$$\chi_{\text{raw}} = \frac{[P_H(c) - P]^2}{8\pi E_\infty^* W c^3} \tag{17}$$

The quantity $\chi_{\text{raw}}$ is obtained by treating the elliptical contact as an area-equivalent circular contact. It provides a global measure of the apparent separation energy but is not the exact local energy-release rate along the elliptical perimeter.

Similarly, the rate of change of the area-equivalent radius is related to the rates of change of the two semi-axes through

$$\frac{dc}{dt} = \frac{1}{2}\sqrt{\frac{b}{a}}\frac{da}{dt} + \frac{1}{2}\sqrt{\frac{a}{b}}\frac{db}{dt} \tag{18}$$

Therefore, $-dc/dt$ represents a global area-recession rate rather than the crack speed at a specific point on the contact edge. The reduced elliptical model retains the functional form of the circular Johnson law because the underlying mechanism remains viscoelastic crack opening at the PDMS interface. Moreover, the approximately self-similar contact evolution indicates that the dominant unloading response can be represented through a single scalar contact coordinate. The circular constitutive form is therefore used as the lowest-order physically motivated rate law, while two effective corrections account for the projection of the non-axisymmetric problem onto $c$.

First, a dimensionless mapping factor $\Gamma_{\text{map}}$ is introduced to relate the circularized apparent energy ratio to the effective driving ratio entering the reduced evolution law:

$$\chi_{\text{map}} = \frac{\chi_{\text{raw}}}{\Gamma_{\text{map}}} \tag{19}$$

Second, the local circular crack-speed scale $C_v$ is replaced by an effective global recession-rate scale $C_{v,\text{eff}}$. The area-equivalent contact evolution is then represented as

$$-\frac{dc}{dt} = C_{v,\text{eff}} \frac{\chi_{\text{map}}}{\ln\left[\frac{1-k}{1-1/\chi_{\text{map}}}\right]} \tag{20}$$

The modulus ratio $k$ is fixed at the value obtained from the circular-contact analysis. The parameters $\Gamma_{\text{map}}$ and $C_{v,\text{eff}}$ should not be interpreted as independent intrinsic material properties. The former corrects the global apparent driving ratio for the loss of local crack-front information, while the latter relates the rate of change of the area-equivalent coordinate to the underlying nonuniform crack-front motion.

Individual crossed-cylinder fits were first used to assess whether these effective quantities varied systematically with crossing angle or preload. Over the range $50° \leq \phi \leq 80°$, both quantities varied only weakly, motivating a second analysis in which one common pair of parameters was applied to all angles and preloads without dataset-specific adjustment. A nested model with $\Gamma_{\text{map}} = 1$ and one globally optimized $C_{v,\text{eff}}$ was also evaluated to determine whether the mapping factor could be absorbed into a simple change in rate scale.

Because increasing $\Gamma_{\text{map}}$ reduces the effective driving ratio $\chi_{\text{map}}$, whereas increasing $C_{v,\text{eff}}$ increases the overall recession-rate scale, the two parameters need not be independently identifiable from the measured $c(t)$ histories. Their joint dependence was therefore examined through the global objective landscape in the $(\Gamma_{\text{map}}, C_{v,\text{eff}})$ plane. The reduced model is applied from load reversal to the maximum tensile load, over which the contact aspect ratio remains approximately constant and the evolution is dominated by the recession of the total contact area. Beyond this point, the contact may undergo increasingly nonuniform crack-front motion, loss of self-similarity, and terminal snap-off, none of which can be represented completely by the single coordinate $c$. The formulation should therefore be viewed as a reduced-order description of the stable viscoelastic recession of moderately elliptical contacts, rather than as a complete local theory of elliptical crack-front propagation.

## 4. Results and Discussion

### *4.1. Circular contacts: geometry-independent formation and dissipative detachment*

The circular-contact configurations provide a reference against which the behavior of the non-circular crossed-cylinder contacts can be assessed. Three nominally different configurations were considered (Fig. 3): cylinder-on-cylinder (C-o-C, 90º (identical cylinders crossed at 90º)), sphere-on-flat (S-o-F), and sphere-on-sphere (S-o-S). Although these specimens differ in their global geometry, the local undeformed gap is axisymmetric in each case and can be described using an appropriate effective radius of curvature. For the crossed-cylinder contact at 90°, $R_{\text{eff}} = R$; for sphere-on-flat, $R_{\text{eff}} = R$; and for two identical spheres, $R_{\text{eff}} = R/2$. The three configurations should therefore obey the same circular JKR relation after the corresponding curvature reduction.

The measured contact-radius–load trajectories exhibit substantial hysteresis between loading and unloading, Fig. 3(a). During loading, the contact radius increases continuously with compressive force for all three configurations. At a given load, the contact sizes differ because the effective radii are different; the sphere-on-sphere configuration, having the smallest $R_{\text{eff}}$, produces the smallest contact. The loading trajectories are otherwise qualitatively similar, whereas the unloading branches show an initial region in which the load decreases considerably with only a small change in contact radius, followed by progressive contact recession and eventual snap-off.

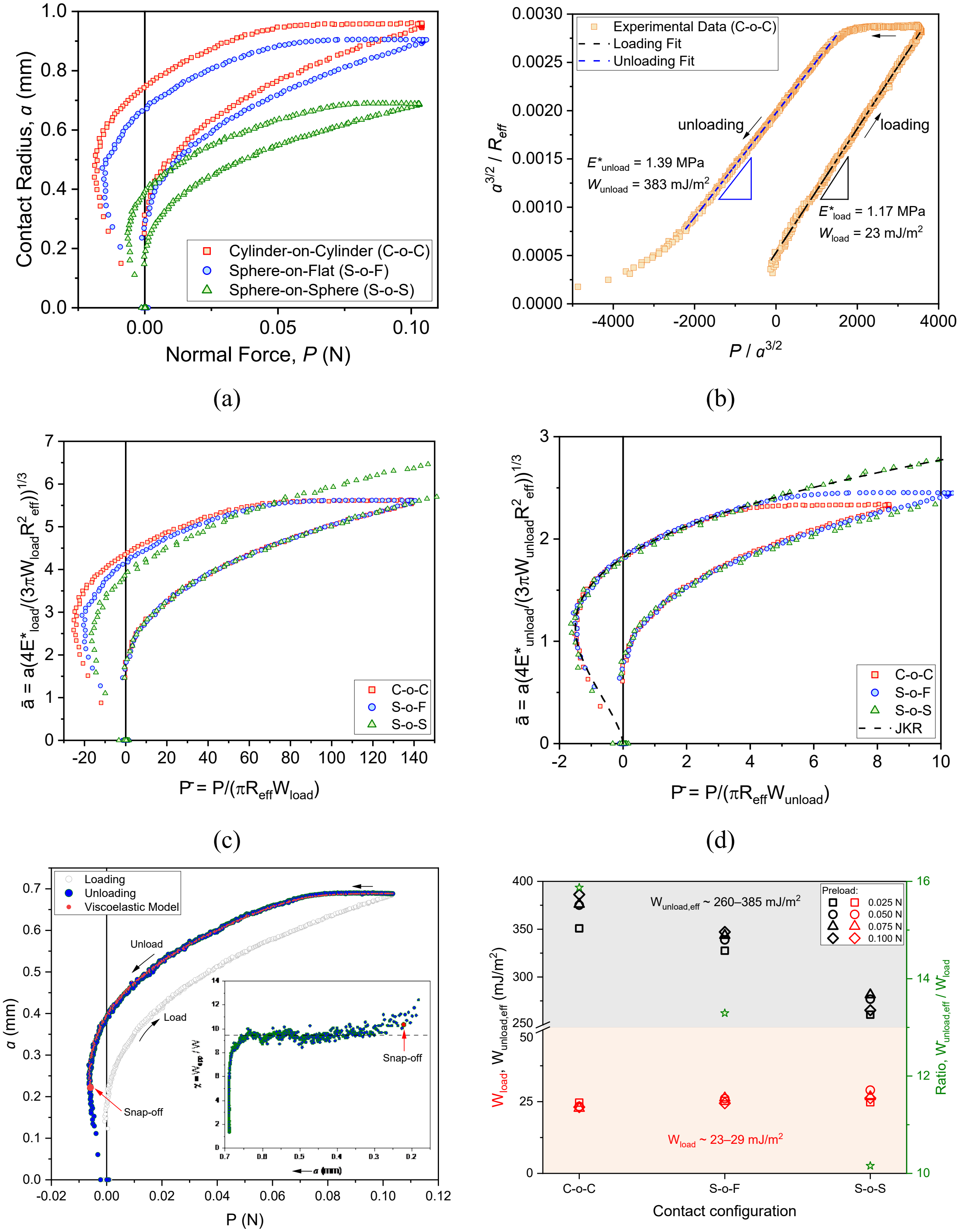

(e) (f)

**Figure 3. (a)** Measured contact-radius–load trajectories for cylinder-on-cylinder (C-o-C) (specifically 90° crossing angle in this figure), sphere-on-flat (S-o-F), and sphere-on-sphere (S-o-S) contacts, showing pronounced hysteresis between the loading and unloading branches. **(b)** Representative C-o-C, 90º response together with linearized loading and unloading fits used to determine the corresponding effective modulus and work of adhesion. **(c)** Comparison of the three contact geometries after normalization using the loading parameters, showing close agreement along the compressive loading branch but deviation during unloading. **(d)** Normalization using the unloading parameters, for which the receding-contact trajectories approach the JKR prediction, indicated by the dashed curve, and exhibit an approximately geometry-independent effective collapse over most of the unloading branch. **(e)** Representative comparison between the experimental unloading trajectory and the viscoelastic contact model; the associated evolution of the apparent-energy ratio, $\chi = W_{\text{app}}/w$, is also shown. **(f)** Loading and effective unloading work of adhesion obtained at different preloads for the three circular contact geometries. The loading work remains approximately $23\text{–}29\ \text{mJ m}^{-2}$, whereas the effective unloading work increases to approximately $280\text{-}385\ \text{mJ m}^{-2}$. The corresponding ratios $W_{\text{unload,eff}}/W_{\text{load}}$ (green star markers) are shown on the secondary axis and is approximately 10-16, demonstrating the substantial viscoelastic enhancement of the apparent separation energy during detachment depending on the geometry.

The loading and unloading branches were first analyzed independently using the linearized JKR representation, Eqn. 8, with $c = a$. A representative 90°crossed-cylinder result is shown in Fig. 3(b). The loading branch is nearly linear and gives approximately $E^*_{\text{load}} \simeq 1.17$ MPa, $W_{\text{load}} \simeq 23\ \text{mJ m}^{-2}$. The extracted loading work of adhesion is consistent with previously reported values for PDMS adhesive contacts. Waters and Guduru [24] obtained $W_{\text{load}}$=25 mJ m$^{-2}$ for a PDMS–glass interface, while the present PDMS-PDMS circular contacts give 25 ± 2 mJ m$^{-2}$. The close agreement is noteworthy because the opposing material-pair differs, suggesting that the effective energy associated with contact formation is comparatively insensitive to the distinction between PDMS-glass and PDMS-PDMS interfaces under the respective low-rate loading conditions.

The receding portion of the unloading branch, Fig. 3b, can also be represented by an approximately linear relation, but the fitted parameters are different: $E^*_{\text{unload}} \simeq 1.39$ MPa, $W_{\text{unload,eff}} \simeq 383\ \text{mJ m}^{-2}$. The modest difference between the fitted loading and unloading moduli may reflect rate dependence, differences between advancing and receding contact states, and the effective nature of the elastic fit applied to a viscoelastic trajectory. The much larger change in the energy scale is more significant. The unloading value should not be interpreted as the equilibrium thermodynamic work of adhesion; it is an effective separation energy incorporating intrinsic adhesion and dissipative resistance to contact-edge recession.

The equivalence of the three circular configurations during contact formation becomes evident when the data are normalized using the loading parameters, Fig. 3(c). The dimensionless variables are:

$$\bar{P}_{\mathrm{load}} = \frac{P}{\pi R_{\mathrm{eff}} W_{\mathrm{load}}}, \quad \bar{a}_{\mathrm{load}} = a\left(\frac{4E^{*}_{\mathrm{load}}}{3\pi W_{\mathrm{load}} R^{2}_{\mathrm{eff}}}\right)^{1/3} \tag{21}$$

The compressive loading branches for C-o-C, S-o-F, and S-o-S collapse closely despite the different specimen geometries. This collapse demonstrates that the local contact response is governed primarily by $R_{\mathrm{eff}}$, $E^{*}_{\mathrm{load}}$, and $W_{\mathrm{load}}$, rather than by the global shape of the specimens. It also confirms that equal-radius cylinders crossed at $90°$ behave, to a good approximation, as a conventional circular JKR contact.

The unloading branches do not collapse when normalized using the loading parameters. In particular, the maximum tensile load occurs at $\bar{P}_{\mathrm{load}} \approx -10$ to $-25$, rather than at the equilibrium JKR value of $-3/2$. In the subsequent discussions below, "pull-off load" has been used, and it refers to the maximum tensile load attained during unloading, not to the terminal snap-off point at which the remaining contact vanishes. This large displacement of the pull-off point directly demonstrates that separation is governed by an energy scale much larger than $W_{\mathrm{load}}$. Using the JKR pull-off relation as an effective estimate,

$$\bar{P}^{(\mathrm{load})}_{\mathrm{pull-off}} \approx -\frac{3}{2}\frac{W_{\mathrm{unload,eff}}}{W_{\mathrm{load}}} \tag{22}$$

a normalized pull-off load of $-20$, for example, corresponds to an effective energy amplification of approximately $13.3$. Thus, the loading-normalized representation provides a direct graphical measure of adhesion hysteresis. The degree of hysteresis also differs among the three circular configurations. At comparable experimental conditions, the magnitude of the normalized tensile load follows approximately

$$\left|\bar{P}_{\mathrm{pull-off}}\right|_{\mathrm{C-o-C}} > \left|\bar{P}_{\mathrm{pull-off}}\right|_{\mathrm{S-o-F}} > \left|\bar{P}_{\mathrm{pull-off}}\right|_{\mathrm{S-o-S}}$$

Because the material formulation, nominal displacement rate, preload, dwell time, and environmental conditions were held fixed, this ordering suggests that the global specimen geometry affects the amount of energy dissipated during detachment, even when the loading contact mechanics is rendered equivalent through $R_{\mathrm{eff}}$. Possible contributions include differences in the volume of material participating in deformation, confinement beneath the contact, specimen compliance, stored bulk viscoelastic energy, and the crack-edge velocity history. The result therefore distinguishes a largely universal contact-formation response from a geometry-dependent separation response.

When the trajectories are instead normalized using the effective unloading parameters, the receding portions of the three contacts approach the JKR prediction over their common normalized-load range, Fig. 3(d). The tensile portions exhibit particularly close agreement. This region is more physically informative than the high-compression region because it corresponds directly to contact-edge recession and fracture-like separation. The collapse indicates that, over a substantial portion of the recession regime, the dissipative response can be represented approximately by replacing the equilibrium adhesion with a constant effective separation energy, $W_{\mathrm{load}} \longrightarrow W_{\mathrm{unload,eff}}$. This does not imply that the underlying process is elastic. Rather, it indicates that the rate-dependent apparent fracture energy varies sufficiently weakly over much of the measured recession interval that a single effective value provides an accurate global representation.

Deviations from the effective JKR curve occur immediately after load reversal. In this regime, the load decreases while the contact radius remains nearly unchanged. Such behavior cannot be captured by a quasi-static elastic JKR relation because the contact edge has not yet entered a steady recession regime. The unloading response can therefore be divided into two stages: an initial delayed-recession or approximately constant-area regime, followed by progressive edge recession that approaches an effective JKR trajectory. A comparable structure is visible in the normalized unloading data reported by Waters and Guduru for a PDMS-glass contact [24]. Although they emphasized that the complete unloading branch is nonlinear and cannot be described by a single constant work of adhesion, their Fig. 5, contains an extended intermediate portion that is approximately linear before stronger terminal curvature develops. Their observation is therefore consistent with the present two-stage interpretation: the entire withdrawal process is not elastic

JKR, but the principal receding-contact stage can be represented over a finite interval by an effective linearized JKR relation.

The physical origin of the initial delayed-recession behavior was examined using Johnson's viscoelastic adhesive-contact model, Fig. 3(e). The contact state at the end of the 10 s dwell was used as the initial condition for the unloading calculation, and the model was applied only to the unloading branch. The reconstructed trajectory reproduces both the initial weak change in contact radius and the subsequent progressive recession up to the vicinity of snap-off. This agreement shows that the nearly horizontal portion of the experimental trajectory is a consequence of viscoelastic delay.

The corresponding apparent-energy ratio, $\chi = W_{\mathrm{app}}/\mathrm{W}$, rises rapidly after unloading begins and then remains approximately constant over a substantial portion of the receding-contact branch. The quasi-plateau indicates that viscoelastic dissipation produces an approximately constant amplification of the intrinsic work of adhesion during much of the recession process. This explains why a single elevated $W_{\mathrm{unload,eff}}$ can reproduce the experimental $a$-$P$ trajectory even though the underlying crack-opening response is viscoelastic. The subsequent increase in $\chi$ near final detachment marks the approach to terminal instability and shows that the constant-energy representation becomes less accurate in the immediate vicinity of snap-off. Applying the same interval-based interpretation to the approximately linear receding portion of the Waters–Guduru data gives an estimated effective unloading separation energy of roughly 217 $\mathrm{mJm^{-2}}$ for the PDMS-glass interface. This value is an estimate obtained from their published linearized data. The value is comparable in magnitude to the present PDMS–PDMS values, although somewhat lower than the minimum value of approximately 260 $\mathrm{mJm^{-2}}$ measured here. The difference may reflect interface chemistry, but it may also arise from differences in unloading rate, specimen compliance, curing history, and the selected fitting interval. The comparison therefore provides independent support for the magnitude of the present effective separation energies without implying that the numerical difference is controlled solely by the material pair.

The fitted energy scales for the different circular contacts are summarized in Fig. 3(f). Across the investigated preloads, the loading work of adhesion remains tightly bound, $W_{\mathrm{load}} \simeq 25 \pm 2\ \mathrm{mJ\ m^{-2}}$, and exhibits only weak dependence on contact configuration and preload. By contrast, the effective unloading energy is broad, $W_{\mathrm{unload,eff}} \simeq 327 \pm 45\ \mathrm{mJ\ m^{-2}}$, corresponding to $W_{\mathrm{unload,eff}}/W_{\mathrm{load}} \simeq 10\text{-}16$. The loading response therefore reflects a comparatively robust contact-formation energy that is nearly independent of circular contact configuration and, when compared with Waters and Guduru, appears similar even across PDMS-PDMS and PDMS-glass interfaces. Detachment, in contrast, is governed by a substantially larger and history-dependent effective separation energy.

### *4.2. Evolution of non-circular contacts*

Having established the response of circular contacts, we next consider the non-orthogonal crossed-cylinder configurations. Decreasing the cylinder crossing angle below 90° produces unequal principal radii of curvature and transforms the contact footprint from circular to elliptical. Representative images for a crossing angle of 60° are shown in Fig. 4(a). During loading, the contact initially forms (snap-in) as a relatively small ellipse and expands progressively as the compressive load increases. Both principal dimensions increase, while the orientation of the major axis remains aligned with the direction imposed by the crossed-cylinder geometry. During unloading, the footprint retraces a distinct path: the contact remains comparatively large immediately after load reversal and subsequently recedes as the load becomes tensile, before undergoing final snap-off.

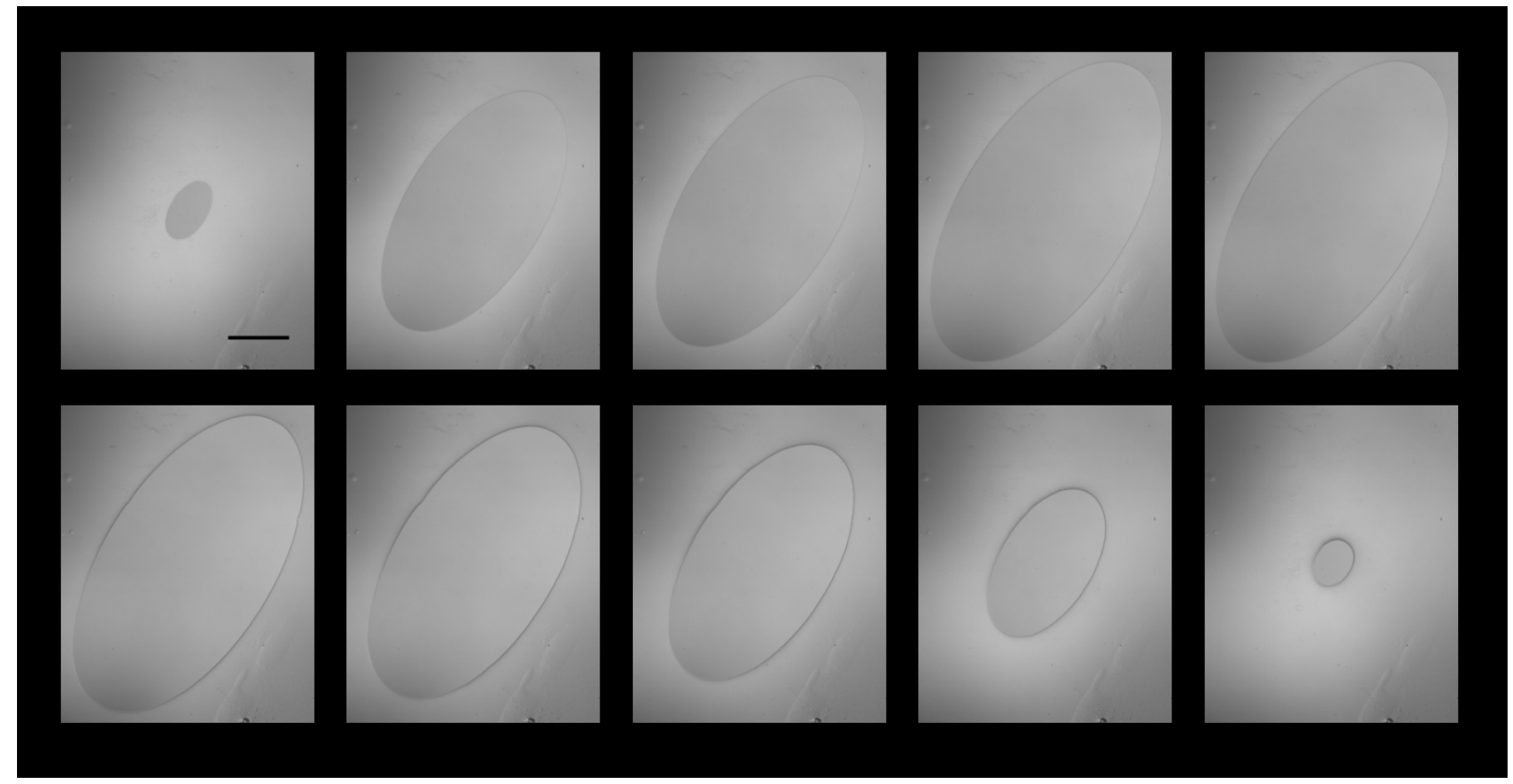

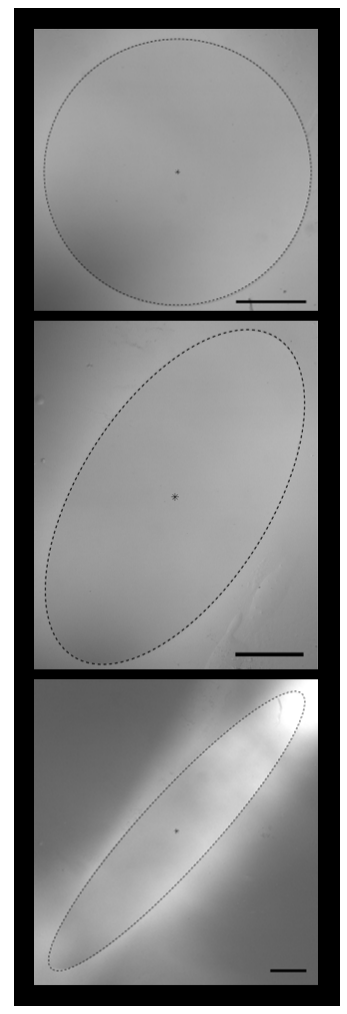

(a) (b)

**Fig. 4. (a)** Representative evolution of the contact footprint during loading (top row) and unloading (bottom row) for the cylinder-on-cylinder (C-o-C) configuration at a crossing angle of 60°(scale bar: 0.5 mm). **(b)** Contact perimeters extracted by edge detection and the corresponding best-fit ellipses used to determine the major and minor semi-axes. The recorded images were first processed in MATLAB using Gaussian filtering and automatic contrast adjustment. The contact boundary was then identified using Canny edge detection, and the resulting perimeter was fitted with a circle or ellipse, as appropriate, to determine the contact dimensions.

The contact perimeter remains well approximated by an ellipse with two semi-axes, $a$ and $b$, over most of the loading and unloading cycle. Fig. 4(b) shows best-fit ellipses applied to the extracted perimeters. The resulting semi-major and semi-minor axes were used to calculate the ellipse ratio, $g = b/a$, and the area-equivalent contact radius, $c = \sqrt{ab}$.

The ellipse fitting is intended to provide a global representation of the footprint. Small local deviations of the contact boundary from the fitted ellipse may occur because of surface heterogeneity, optical noise, imperfect alignment, or nonuniform crack-front motion, particularly during the final stages of detachment. Nevertheless, the fitted ellipses capture the dominant contact dimensions and preserve the measured contact area through $A = \pi ab = \pi c^2$. The suitability of $c$ as a scalar contact-size variable depends not only on the quality of the ellipse fit, but also on whether the footprint changes mainly in size rather than continually changing shape. This question is examined directly through the evolution of $a$, $b$, and $g$ in Fig. 5(a,b).

### *4.3. Approximate self-similarity of elliptical contacts*

Figure 5(a) shows the measured semi-major and semi-minor axes during loading and unloading for different crossing angles. As expected, decreasing the crossing angle increases the anisotropy of the contact. At 80°, the two dimensions remain relatively close, whereas at 30°, the major axis becomes several times larger than the minor axis. Both dimensions increase continuously during loading, but their absolute rates of growth differ according to the imposed curvature anisotropy.

Despite the large differences in contact shape between angles, the ratio $g = b/a$ exhibits a comparatively simple evolution, Fig. 5(b). During the earliest stage of contact formation, $g$ changes rapidly as the initially small footprint adjusts to the local crossed-cylinder geometry. After this initial shape-adjustment regime, $g$ approaches an angle-dependent plateau. The plateau value decreases systematically with crossing angle, from $g \approx 1$ for the circular 90°contact to approximately 0.2 for the most elongated 30°contact. For a given crossing angle, $g$ then varies only weakly over most of the compressive loading branch. Thus, $b \approx g_{\phi}a,$ where $g_{\phi}$ is approximately constant for a given crossing angle. The contact therefore grows approximately as a family of geometrically similar ellipses. Under this condition, $A = \pi ab = \pi g_{\phi} a^2,$ and the dominant evolution of the contact is represented by a change in overall scale rather than a substantial change in shape.

The same approximate self-similarity persists through much of unloading. Following load reversal, the contact dimensions initially change only weakly, after which both semi-axes decrease as the contact recedes. Importantly, $g$ remains approximately constant through the compressive unloading range and into the tensile regime up to approximately the maximum tensile load. Changes in the contact footprint over this mechanically important interval are therefore dominated by the loss of contact area rather than by a large redistribution between the two principal axes. The approximately constant aspect ratio provides the experimental basis for using $c = \sqrt{ab}$ as a single global contact coordinate. The area-equivalent radius is always an exact measure of the footprint area, regardless of whether $g$ is constant. However, the weak variation of $g$ gives the stronger result that $c$ also captures most of the contact evolution without discarding a major independent shape change. In this sense, the contact behaves approximately as a one-parameter family of ellipses over the principal loading and unloading intervals.

Noticeable deviations from the plateau occur near the inception of contact, and during terminal detachment. These deviations identify the regimes in which the local crack-front evolution becomes increasingly nonuniform and a scalar area-equivalent description is least likely to represent the complete mechanics. The area-equivalent analysis below is therefore interpreted as a global contact representation rather than as an assertion that the local energy-release rate is uniform around the elliptical perimeter.

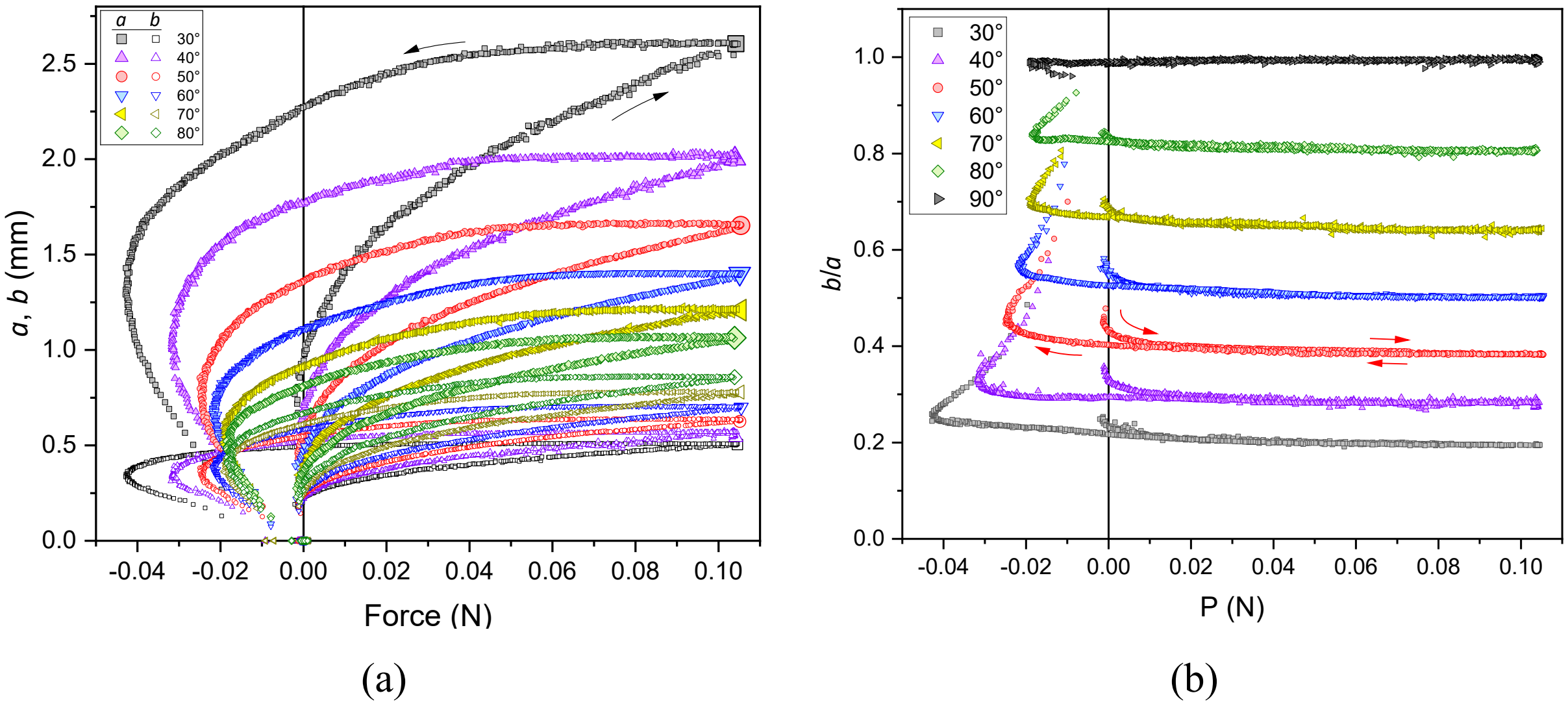

**Fig. 5. (a)** Measured semi-major and semi-minor axes, $a$ and $b$, respectively, as a function of normal load during loading and unloading for various crossing angles at one preload level. **(b)** evolution of the contact aspect ratio $g = b/a$ during loading and unloading for different cylinder crossing angles. After an initial shape-adjustment regime, $g$ approaches an angle-dependent plateau and remains approximately constant throughout the compressive loading range and over most of unloading, including the tensile regime up to the maximum tensile load.

### *4.4. Area-equivalent representation of contact formation*

The approximately self-similar loading response motivates testing whether the non-circular contacts can be represented using the area-equivalent JKR variables $c$ and $R_{\text{eff}}$. Figure 6(a) shows the corresponding linearized trajectories for representative crossing angles of 30°, 50°, and 80°. The loading branches are approximately linear over the fitted range despite the substantial difference in ellipticity between these contacts. The linearity indicates that the loading response can be written in the form of Eqn. 8. Thus, once the measured area and the geometric-mean curvature are used, a circular JKR-type relation gives a compact description of contact formation even when the actual contact edge is elliptical.

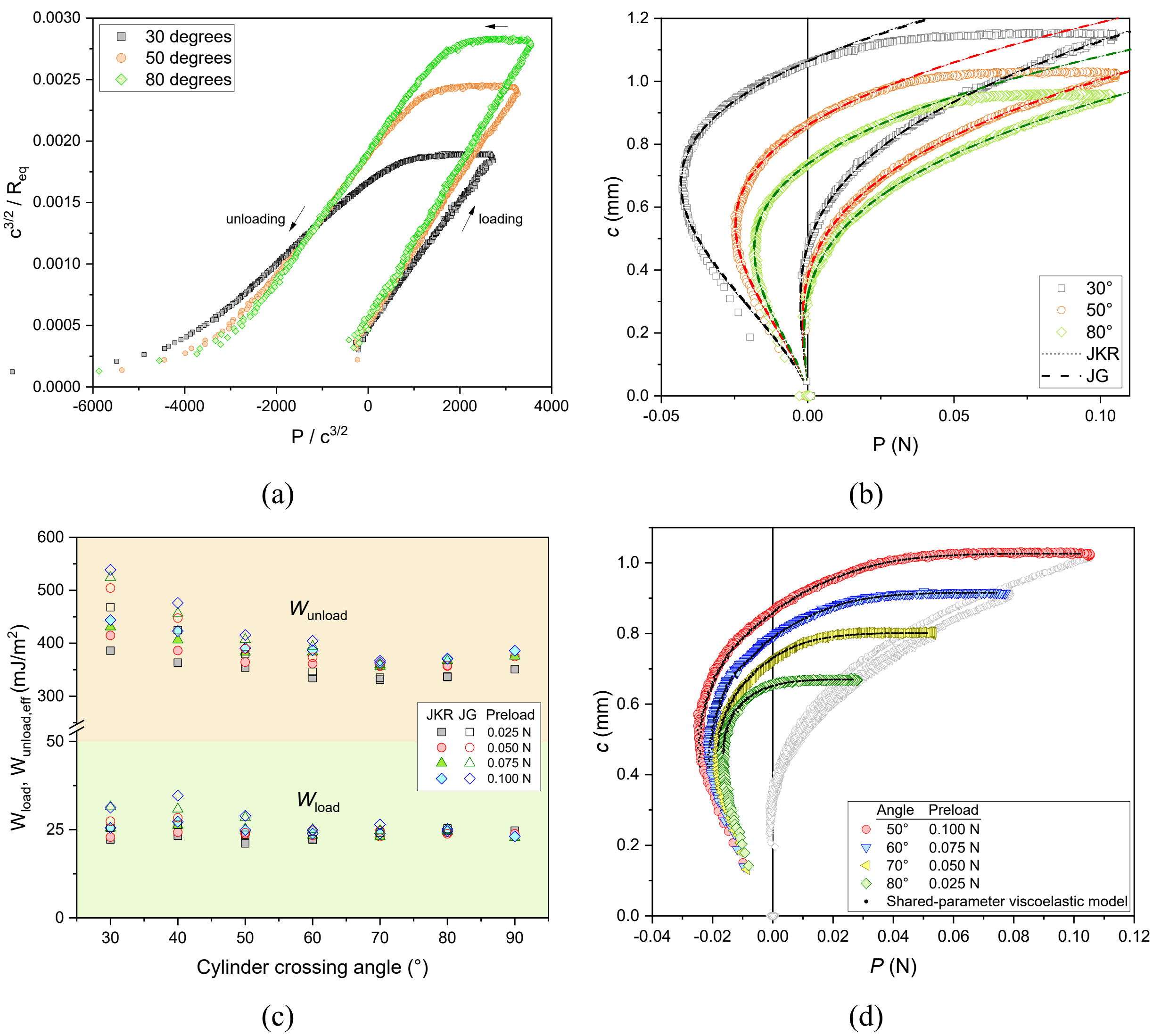

**Fig. 6. (a)** Linearized loading and unloading data used to determine the corresponding effective modulus and work of adhesion for crossing angles of 30°, 50°, and 80°. **(b)** Representative loading and unloading *c*–*P* responses for the same crossing angles, together with the corresponding area-equivalent JKR and Johnson–Greenwood fits. **(c)** Summary of the loading and effective unloading work of adhesion obtained from the JKR and Johnson–Greenwood analyses for the different crossing angles and preload levels. Marker shape denotes preload, while marker color and fill distinguish the fitted model and loading or unloading branch. **(d)** Representative crossed-cylinder unloading responses compared with the shared-parameter reduced viscoelastic model. The same $\Gamma_{\text{map}}$ and $C_{v,\text{eff}}$ are used for all cases; gray symbols show the corresponding loading branches.

Representative *c*-*P* trajectories and fits are shown in Fig. 6(b). During loading, the area-equivalent JKR and Johnson–Greenwood curves are nearly indistinguishable over most of the measured range. Both closely reproduce the experimental contact growth for the representative 30°, 50°, and 80°contacts. The agreement is significant because Johnson–Greenwood explicitly incorporates the elliptical geometry, whereas the area-equivalent JKR relation retains only the scalar variables $c$ and $R_e$. The comparable quality of the two fits indicates that detailed pointwise changes in the ellipse ratio contribute relatively little to the global loading trajectory once the footprint has entered the approximately self-similar regime. In other words, contact formation is governed primarily by the increase in total area, while the angle-dependent shape is absorbed effectively through the geometric-mean curvature and the measured area-equivalent radius.

The fitted loading energies are summarized in the lower portion of Fig. 6(c). Across the investigated crossing angles and preloads, the area-equivalent JKR analysis gives $W_{\mathrm{load}} \approx 24 \pm 1\ \mathrm{mJ\ m^{-2}}$. The values obtained from the Johnson–Greenwood fits are comparable, $W_{\mathrm{load}} \approx 26 \pm 3\ \mathrm{mJ\ m^{-2}}$, though the deviation is slightly higher. Neither analysis shows a strong systematic dependence of $W_{\mathrm{load}}$ on crossing angle or preload. This is consistent with the expectation that the energy associated with formation of the same PDMS–PDMS interface should not change merely because the contact footprint is made more elliptical.

The agreement between the two models is also important methodologically. A good area-equivalent JKR fit alone could be regarded as an empirical consequence of choosing $c$. The independent agreement with the explicit elliptical-contact model demonstrates that the nearly angle-independent loading energy is not simply an artifact of circularizing the contact area. Rather, over the present range of geometries, the area-equivalent variables retain the dominant information required to describe the global loading response.

The combined results from Fig. 6(a-c) therefore establish the principal loading result of this study: after an initial shape-adjustment regime, crossed-cylinder contacts evolve approximately self-similarly, and their global contact formation can be represented by an area-equivalent JKR relation that yields a nearly geometry- and preload-independent loading work of adhesion.

### 4.5. *Geometry-amplified detachment of crossed-cylinder contacts*

The unloading response differs fundamentally from loading. Figure 6(a) shows that immediately after load reversal, the linearized trajectory does not follow the fitted loading line or a single equilibrium unloading line. Instead, the load decreases substantially while the contact area changes only weakly. This is the elliptical-contact counterpart of the delayed-recession regime identified for the circular contacts in Section 4.1.

Neither the area-equivalent JKR model nor the elastic Johnson–Greenwood model captures this initial portion of unloading, Fig. 6(b). This limitation is expected because both relations assume that the contact edge responds according to an instantaneous elastic fracture condition. During the delayed-recession stage, however, the load changes before substantial contact-edge motion occurs. After this transient, both semi-axes begin to decrease and the contact enters a progressive recession regime. Over a substantial portion of this second stage, the transformed unloading data become approximately linear, Fig. 6(a). The decreasing-$c$ branch can therefore be represented using the same global contact variables, but with an effective unloading modulus and a substantially increased effective separation energy, $W_{\mathrm{unload,eff}}$.

As in the circular-contact analysis, this quantity is not interpreted as a thermodynamic work of adhesion. It is an effective energy that incorporates the resistance associated with viscoelastic deformation, interfacial dissipation, and nonuniform recession of the elliptical contact edge.

The area-equivalent JKR and Johnson–Greenwood fits are again similar over much of the progressive recession regime, Fig. 6(b). Their agreement does not imply that either elastic model describes the complete unloading process. Rather, it shows that once the initial delayed-recession interval is excluded, a constant effective energy provides a useful global approximation over the principal decreasing-area branch. Deviations become more pronounced beyond the maximum tensile-load point and close to snap-off, where the contact becomes small and the crack-front motion accelerates.

The fitted energy values in Fig. 6(c) reveal a strong asymmetry between formation and separation. Whereas $W_{\mathrm{load}}$ remains confined to a narrow band, $W_{\mathrm{unload,eff}}$ is more than an order of magnitude larger. Moreover, its dependence on geometry is systematic: the effective unloading energy increases as the crossing angle decreases. The most elongated $30°$ contacts exhibit the largest values, while the nearly circular $80°$and $90°$contacts give lower values.

Preload also affects the unloading response. At a given angle, increasing the maximum compressive preload generally increases $W_{\mathrm{unload,eff}}$, and this preload sensitivity becomes more pronounced as the contact becomes more elongated. In contrast, the loading adhesion remains nearly insensitive to preload. The contact therefore retains little memory of its maximum size during formation but a strong memory of preload during separation.

This distinction suggests that the additional energy is not an equilibrium interface property. A larger preload creates a larger contact and deforms a greater volume of PDMS before unloading. During subsequent separation, part of the stored and dissipated energy depends on the deformation history. In an elongated contact, the deformation and crack-front recession are also distributed anisotropically: different positions around the perimeter experience different local curvatures, energy-release rates, and recession velocities. The scalar $W_{\mathrm{unload,eff}}$ averages these processes into one global quantity, and its increase with elongation is consistent with progressively stronger geometry-dependent dissipation.

The results therefore identify geometry-amplified detachment. Contact elongation has only a weak effect on the energy extracted during loading, but it substantially increases the effective resistance to separation and its sensitivity to preload. The principal influence of non-circular geometry thus appears during crack opening rather than crack closure.

The influence of preload on pull-off in soft adhesive contacts remains unsettled. Previous studies on nominally circular sphere–flat contacts have reported both negligible and pronounced preload dependence, with the outcome attributed to differences in viscoelastic relaxation, loading history, pulling rate, and finite-size dissipation [26]. In the present experiments, the circular contacts exhibit comparatively weak preload sensitivity, whereas a clear preload dependence emerges as the crossed-cylinder contact becomes increasingly elongated. This suggests that contact geometry may constitute an additional control parameter in determining whether preload history is manifested in the pull-off response. In particular, while compact circular contacts appear relatively insensitive to the maximum preceding load under the present conditions, elongated contacts exhibit progressively stronger history dependence during detachment. This behavior is consistent with the possibility that increasing contact anisotropy alters the volume and spatial distribution of viscoelastic deformation and produces more nonuniform crack-front recession, thereby amplifying the influence of the preceding preload. The present results, therefore, suggest that contact geometry is an additional factor that may help reconcile apparently conflicting reports of preload-dependent pull-off in soft adhesive contacts.

### *4.6. Reduced viscoelastic description of elliptical-contact unloading*

The pronounced hysteresis observed during unloading suggests that the enhanced resistance to separation cannot be interpreted solely in terms of an equilibrium work of adhesion. As shown for the circular contacts, much of this behavior can instead be rationalized through a rate-dependent resistance to contact-edge recession. We therefore examined whether the same viscoelastic framework could provide a reduced description of the non-circular crossed-cylinder contacts. For this purpose, the elliptical footprint was represented by the area-equivalent contact radius, $c$, while the crossed-cylinder geometry was incorporated through the effective radius $R_{\mathrm{eff}}$. The model consequently describes the evolution of the total contact area through a single scalar coordinate and should be regarded as an effective representation of elliptical-contact recession rather than a complete solution for the spatially non-uniform motion of an elliptical crack front.

The material parameters $E_{\infty}^{*}$, $w$, and $k = E_{\infty}^{*}/E_{0}^{*}$ were fixed at the values determined independently from the circular-contact analysis. A direct application of the circular viscoelastic relation to the area-equivalent elliptical contact would correspond to setting the mapping factor $\Gamma_{\mathrm{map}} = 1$. We first tested this limiting case by fixing $\Gamma_{\mathrm{map}} = 1$ and determining the characteristic velocity $C_{v,\mathrm{eff}}$ that best described the crossed-cylinder unloading data. Although adjustment of $C_{v,\mathrm{eff}}$ changed the overall rate of predicted contact recession, the model did not adequately reproduce the measured unloading trajectories (Fig. S2 and Table S1). These results indicate that replacing the elliptical contact simply by $c$ and $R_{\mathrm{eff}}$ is insufficient to transfer the circular crack-driving relation quantitatively, and that an additional mapping of the apparent driving parameter is required.

We next allowed both $\Gamma_{\mathrm{map}}$ and $C_{v,\mathrm{eff}}$ to vary. The quality of the individual fits improved substantially; however, the two fitted parameters exhibited appreciable covariance. Different combinations of $\Gamma_{\mathrm{map}}$ and $C_{v,\mathrm{eff}}$ within a relatively narrow region of parameter space generated similar $c - P$ trajectories and comparable values of the fitting objective. Consequently, the precise values obtained from any single unloading trajectory should not be interpreted independently as uniquely determined physical parameters. This covariance is illustrated further in the Supplementary Information, Fig. S3, through the objective landscape in the $(\Gamma_{\mathrm{map}}, C_{v,\mathrm{eff}})$ parameter space.

We therefore examined whether the locations of the best-fit parameter pairs varied systematically with crossing angle or preload. Each dataset over $50° \leq \phi \leq 80°$ was first fitted independently with both $\Gamma_{\mathrm{map}}$ and $C_{v,\mathrm{eff}}$ adjustable. Despite the covariance within individual fits, the best-fit values obtained across the dataset ensemble showed no systematic dependence on either angle or preload and remained within approximately 10% of their respective mean values. This limited variation motivated the use of a common parameter pair. The arithmetic means of the independently fitted values, $\Gamma_{\mathrm{map}} = 1.69$, $C_{v,\mathrm{eff}} = 1.16 \times 10^{-7}$ m s$^{-1}$, were therefore fixed and subsequently applied to all crossed-cylinder datasets over the range without further dataset-specific adjustment. All parameters used are provided in Table S2. Thus, the curves shown in Fig. 6(d) are not individual best fits, but predictions obtained using a single shared parameter pair derived from the ensemble of independent fits.

Figure 6(d) shows representative comparisons for $50°/0.100$N, $60°/0.075$N, $70°/0.050$N, and $80°/0.025$N, while results for all investigated angle-preload combinations are provided in the Supplementary Information, Fig. S4. Despite the substantial differences in contact size, ellipticity, and preload, the common-parameter model captures the principal features of the unloading trajectories, including the initially weak recession of the contact area followed by progressively faster reduction in $c$ as the tensile load develops. The ability of the mean parameter values to reproduce these different trajectories without dataset-specific adjustment of either $\Gamma_{\text{map}}$ or $C_{v,\text{eff}}$ is considerably more restrictive than fitting each curve independently and suggests that the different experiments are governed by a common underlying rate-dependent separation mechanism.

The interpretation of the two shared parameters should nevertheless be distinguished. $C_{v,\text{eff}}$ sets the characteristic rate scale for evolution of the area-equivalent contact coordinate and should therefore be regarded as an effective recession-rate parameter rather than an intrinsic material constant. Its weak variation across the independently fitted datasets and the success of a common value indicate that a similar characteristic rate scale governs the unloading response over the investigated range of angles and preloads. In contrast, $\Gamma_{\text{map}}$ is explicitly a geometric mapping parameter introduced to account for the fact that the driving quantity inferred from the area-equivalent representation is not identical to that of an axisymmetric circular crack. The relatively modest correction, $\Gamma_{\text{map}} \approx 1.69$, and its successful use over $50° - 80°$ indicate that the combination of $c$ and $R_{\text{eff}}$ captures much, although not all, of the geometric influence in this range. Application of the common $\Gamma_{\text{map}}$ and $C_{v,\text{eff}}$ values to the lower-angle contacts, Fig. S5, provides an additional test of the reduced model outside the range used to establish the shared parameters. The prediction remains quantitatively close to experiment at $\phi = 40°$, with a preload-averaged NRMSE of approximately 5.7% (Table S3). At $\phi = 30°$, the NRMSE increases to approximately 8.8%, indicating the onset of systematic deviations for highly elongated contacts. In comparison, the corresponding errors over $50° \leq \phi \leq 80°$ are predominantly 2–4%. This progressive loss of accuracy at low crossing angles is consistent with the limitations of representing an increasingly anisotropic crack front through the single area-equivalent coordinate $c$.

The reduced description is expected to become progressively less accurate as the contact approaches terminal detachment. During most of unloading, the approximately constant aspect ratio indicates near-self-similar shrinkage of the elliptical footprint, providing a basis for representing its evolution through the single coordinate $c$. Close to pull-off, however, the aspect ratio changes more rapidly and the motion of the crack front need no longer remain spatially uniform. In this regime, the scalar area-equivalent description cannot distinguish recession along the major and minor axes and may produce nonphysical behavior beyond the principal unloading branch. For this reason, the model comparison in Fig. 6(d) is restricted to the branch extending to the maximum tensile load, rather than interpreting the subsequent terminal collapse as a quantitative prediction of local snap-off. The complete experimental trajectory is retained to show the eventual instability.

The viscoelastic description also clarifies the meaning of the much larger effective separation energies obtained by fitting the unloading curves with an elastic JKR-type relation. In the elastic representation, the entire dissipative unloading response is condensed into an elevated $W_{\mathrm{unload,eff}}$. In the present formulation, the enhanced separation resistance instead emerges dynamically because the contact-edge recession rate depends on the instantaneous crack-driving state. The effective elastic and viscoelastic descriptions should therefore be viewed as complementary. $W_{\mathrm{unload,eff}}$ provides a convenient global measure of the resistance encountered during unloading, whereas the reduced viscoelastic model provides a physically motivated rate-dependent interpretation of how the enhanced apparent separation resistance develops during contact recession. The ability of a single shared parameter set to describe multiple angles and preloads further supports the existence of a common rate-dependent separation process over the moderately elliptical regime, with contact geometry governing how this response is expressed through the evolving non-circular footprint.

## 5. Conclusions

This study examined the formation and detachment of soft adhesive contacts across circular and non-circular geometries using PDMS sphere-on-flat, sphere-on-sphere, and crossed-cylinder configurations. By varying the cylinder crossing angle from 30° to 90°, the contact footprint was changed continuously from highly elongated elliptical to nearly circular while retaining the same material pair and experimental conditions. This geometry therefore provided a controlled means of distinguishing the role of contact shape during formation from its influence during detachment.

During loading, the elliptical contacts evolved approximately self-similarly after an initial shape-adjustment regime, with the aspect ratio $g = b/a$ approaching an angle-dependent plateau. This behavior motivated the use of the area-equivalent radius $c = \sqrt{ab}$, together with the geometric-mean curvature radius $R_e = \sqrt{R_1 R_2}$. With these substitutions, the loading trajectories were described well by an area-equivalent JKR representation. The extracted loading work of adhesion remained within a relatively narrow range of approximately $23 - 29 \text{ mJ m}^{-2}$ and showed only weak dependence on crossing angle and preload. Comparable values obtained from the Johnson-Greenwood analysis further indicate that this result is not simply an artifact of neglecting contact ellipticity. Thus, over the investigated range, contact formation is governed predominantly by the evolution of total contact area despite the non-axisymmetric crack-front geometry.

The detachment response was markedly different. Immediately after load reversal, the contact dimensions changed only weakly while the applied force decreased, producing a delayed-recession or nearly constant-area regime. Once recession commenced, much of the unloading trajectory could be represented by an effective JKR-like relation, but only by using an effective separation energy substantially larger than the loading work of adhesion. For the circular contacts, $W_{\text{unload,eff}}$ was approximately $260 - 385 \text{ mJ m}^{-2}$, corresponding to an enhancement of roughly one order of magnitude relative to $W_{\text{load}}$. The circular-contact viscoelastic model reproduced both the initial delayed recession and the subsequent contact-edge recession, providing a mechanistic basis for the elevated effective separation energy inferred from the elastic unloading fits.

For the crossed-cylinder contacts, unloading became increasingly sensitive to preload as the contact became more elongated. Whereas the loading adhesion remained only weakly dependent on geometry, both the effective unloading energy and pull-off response increased with decreasing crossing angle and increasing preload. The influence of geometry therefore emerges primarily during separation rather than during contact formation. Importantly, the non-circular unloading response could also be described using the reduced area-equivalent viscoelastic formulation. Independent fits over $50° \leq \phi \leq 80°$ produced only modest variations in the fitted mapping factor $\Gamma_{\text{map}}$ and characteristic velocity $C_{v,\text{eff}}$. Their mean values were therefore adopted as a single shared parameter set and, without further dataset-specific adjustment, reproduced the principal unloading features across different crossing angles and preloads. This indicates that the rate-dependent separation response over this range can be represented by a common reduced constitutive law, while the geometric influence is incorporated primarily through the effective curvature, area-equivalent contact size, and the mapping factor $\Gamma_{\text{map}}$.

The area-equivalent radius should nevertheless be interpreted as a global measure of contact size rather than as a local crack-front coordinate. Its success is greatest while the elliptical footprint recedes approximately self-similarly; close to terminal detachment, changes in aspect ratio and nonuniform crack-front motion increasingly limit a scalar description.

In summary, the results establish a clear distinction between contact formation and detachment in soft adhesive interfaces. Contact formation can be reduced, over the investigated geometries and loading conditions, to an area-equivalent JKR description using $c = \sqrt{ab}$ and $R_e = \sqrt{R_1 R_2}$. Detachment, in contrast, retains a strong dependence on preload, loading history, and contact elongation. The elevated $W_{\mathrm{unload,eff}}$ provides a compact measure of this dissipative resistance, while the reduced viscoelastic formulation provides a mechanistic description of its rate dependence. Together, these results provide a practical framework for analyzing non-circular soft adhesive contacts while preserving the distinction between comparatively geometry-insensitive contact formation and geometry-amplified dissipative separation.

## 6. Acknowledgement

The authors acknowledge support from the Anusandhan National Research Foundation (ANRF) through core research grant No. CRG/2023/006357.

## 7. Data and Code Availability

The data that support the findings of this study are available from the corresponding author upon reasonable request. Analysis scripts will be shared for academic use on request.

## 8. Conflict of Interest Statement

The authors declare no competing financial or non-financial interests.

## References

1. Johnson KL, Kendall K, Roberts AD (1971) Surface Energy and the Contact of Elastic Solids. Proc R Soc A Math Phys Eng Sci 324:301–313. https://doi.org/10.1098/rspa.1971.0141
2. Kendall K (1971) The adhesion and surface energy of elastic solids. J Phys D Appl Phys 4:320. https://doi.org/10.1088/0022-3727/4/8/320
3. Chaudhury MK, Whitesides GM (1991) Direct measurement of interfacial interactions between semispherical lenses and flat sheets of poly(dimethylsiloxane) and their chemical derivatives. Langmuir 7:1013–1025. https://doi.org/10.1021/la00053a033
4. Creton C, Ciccotti M (2016) Fracture and adhesion of soft materials: a review. Reports Prog Phys 79:046601. https://doi.org/10.1088/0034-4885/79/4/046601
5. Ciavarella M, Joe J, Papangelo A, Barber JR (2019) The role of adhesion in contact mechanics. J R Soc Interface 16:20180738. https://doi.org/10.1098/rsif.2018.0738
6. Das D, Chasiotis I (2020) Sliding of adhesive nanoscale polymer contacts. J Mech Phys Solids 140:103931. https://doi.org/10.1016/j.jmps.2020.103931
7. Das D, Chasiotis I (2021) Rate dependent adhesion of nanoscale polymer contacts. J Mech Phys Solids 156:104597. https://doi.org/10.1016/j.jmps.2021.104597
8. Derjaguin B., Muller V., Toporov Y. (1975) Effect of contact deformations on the adhesion of particles. J Colloid Interface Sci 53:314–326. https://doi.org/10.1016/0021-9797(75)90018-1
9. Maugis D (1992) Adhesion of spheres: The JKR-DMT transition using a dugdale model. J Colloid Interface Sci 150:243–269. https://doi.org/10.1016/0021-9797(92)90285-T
10. Tabor D, Winterton RHS (1969) The direct measurement of normal and retarded van der Waals forces. Proc R Soc London A Math Phys Sci 312:435–450. https://doi.org/10.1098/rspa.1969.0169
11. Israelachvili JN (2011) Intermolecular and Surface Forces. Elsevier
12. Maeda N, Nianhuan C, Tirrell M, Israelachvili JN (2002) Adhesion and Friction Mechanisms of Polymer-on-Polymer Surfaces. Science (80- ) 297:379–382. https://doi.org/10.1126/science.1072378
13. Zeng H, Maeda N, Chen N, et al (2006) Adhesion and friction of polystyrene surfaces around T g. Macromolecules 39:2350–2363. https://doi.org/10.1021/ma052207o
14. Sümer B, Onal CD, Aksak B, Sitti M (2010) An experimental analysis of elliptical adhesive contact. J Appl Phys 107:. https://doi.org/10.1063/1.3428494
15. Johnson KL, Greenwood JA (2005) An approximate JKR theory for elliptical contacts. J Phys D Appl Phys 38:1042–1046. https://doi.org/10.1088/0022-3727/38/7/012
16. Li Q, Popov VL (2020) A numerical study of JKR-type adhesive contact of ellipsoids. J Phys D Appl Phys 53:335303. https://doi.org/10.1088/1361-6463/ab85e9
17. Giudici A, Vella D, Griffiths I (2025) Mechanics of elliptical JKR-type adhesive contact. J Phys D Appl Phys 58:085301. https://doi.org/10.1088/1361-6463/ad983f
18. Silberzan P, Perutz S, Kramer EJ, Chaudhury MK (1994) Study of the Self-Adhesion Hysteresis of a Siloxane Elastomer Using the JKR Method. Langmuir 10:2466–2470. https://doi.org/10.1021/la00019a073
19. Baek D, Hemthavy P, Saito S, Takahashi K (2017) Evaluation of energy dissipation involving adhesion hysteresis in spherical contact between a glass lens and a PDMS block. Appl Adhes Sci 5:4. https://doi.org/10.1186/s40563-017-0082-z

20. Pickering JP, Van Der Meer DW, Vancso GJ (2001) Effects of contact time, humidity, and surface roughness on the adhesion hysteresis of polydimethylsiloxane. J Adhes Sci Technol 15:1429–1441. https://doi.org/10.1163/156856101753213286
21. Amouroux N, Léger L (2003) Effect of Dangling Chains on Adhesion Hysteresis of Silicone Elastomers, Probed by JKR Test. Langmuir 19:1396–1401. https://doi.org/10.1021/la020680e
22. Petroli A, Petroli M, Romagnoli M, Geoghegan M (2022) Determination of the rate-dependent adhesion of polydimethylsiloxane using an atomic force microscope. Polymer (Guildf) 262:125445. https://doi.org/10.1016/j.polymer.2022.125445
23. Chaudhury MK, Weaver T, Hui CY, Kramer EJ (1996) Adhesive contact of cylindrical lens and a flat sheet. J Appl Phys 80:30–37. https://doi.org/10.1063/1.362819
24. Waters JF, Guduru PR (2010) Mode-mixity-dependent adhesive contact of a sphere on a plane surface. Proc R Soc A Math Phys Eng Sci 466:1303–1325. https://doi.org/10.1098/rspa.2009.0461
25. Greenwood JA, Johnson KL (2006) Oscillatory loading of a viscoelastic adhesive contact. J Colloid Interface Sci 296:284–291. https://doi.org/10.1016/j.jcis.2005.08.069
26. Violano G, Afferrante L (2022) Size effects in adhesive contacts of viscoelastic media. Eur J Mech - A/Solids 96:104665. https://doi.org/10.1016/j.euromechsol.2022.104665